\documentstyle[twocolumn,aps,epsf]{revtex}

\begin{document}


\title{Preparing Topological States of a Bose-Einstein
  Condensate}

\author{J. E. Williams \& M. J. Holland}

\address{JILA and Physics Department, University of Colorado, Boulder,
  CO 80309-0440, USA}


\maketitle
\vskip2pc

{\bf The burgeoning field of Bose-Einstein condensation in dilute
  alkali and hydrogen gases has stimulated a great deal of research
  into the statistical physics of weakly interacting quantum
  degenerate systems$^{\bf 1,2}$. The recent experiments offer the
  possibility for exploring fundamental properties of low temperature
  physics in a very controllable and accessible way. One current goal
  of experimenters in this field is to observe superfluid-like
  behavior in these trapped Bose gases, analogous to persistent
  currents in superfluid liquid helium, which flow without observable
  viscosity, and electric currents in superconductors, which flow
  without observable resistance. These ``super'' properties of
  Bose-condensed systems occur because the macroscopic occupation of a
  quantized mode provides a stabilizing mechanism that inhibits decay
  due to thermal relaxation$^{\bf 3}$. Here we solve the
  time-dependent Gross-Pitaevskii equation of motion of the condensate
  involving two hyperfine atomic states and show how to generate, with
  extremely high fidelity, topological modes such as vortices that
  open the door to the study of superfluidity in these new systems.
  Our approach is inspired by recent experiments investigating a
  trapped condensate with two strongly coupled internal states$^{\bf
    4,5}$. We show how the interplay between the internal and motional
  dynamics can be utilized to prepare the condensate in a variety of
  interesting configurations.}

Since 1995, when Bose-Einstein condensation in a dilute atomic gas was
first observed$^{6-8}$, experimenters have sought a method to create a
vortex in this system. In a typical experiment, around one million
atoms are trapped in a magnetic harmonic potential and cooled below
the critical temperature so that condensation occurs into the lowest
energy quantized mode. In the usual case, this ground state has no
circulation.  One proposed scheme for preparing the condensate in a
vortex mode$^{9-17}$ is to distort the confining potential and
mechanically rotate the trap during the cooling process. In this way,
the lowest energy mode may be engineered to be circulating about the
axis of symmetry. Such an approach is in direct analogy with
experiments on vortices in superfluid helium---the asymmetry of the
harmonic trap for the atomic gas plays the role of surface roughness
of a rotating vessel. Although conceptually this method appears
promising for vortex generation in a trapped gas, so far technical
difficulties have precluded its successful implementation.

Instead of having the system condense into a vortex mode, an
alternative approach is to allow the atoms to condense into the usual
ground state and then dynamically generate the vortex from the
non-rotating condensate. Several theoretical proposals have been made
along these lines which utilize the interaction between the atoms and
a specific laser field consisting of a beam of photons with non-zero
orbital angular momentum$^{18-20}$.

The method we present here makes use of both of the techniques
mentioned---mechanical rotation and the coupling of internal states
using an electromagnetic field. It is motivated by the recent
observations demonstrating the cyclic twisting and untwisting of the
order parameter of a gaseous condensate of rubidium atoms$^4$. These
recent experiments have suggested the possibility for harnessing the
interplay between the internal and motional dynamics in order to
fundamentally alter the topological structure of the order parameter
of the condensate$^5$.

\begin{figure}
\begin{center}\
\epsfysize=50mm
\epsfbox{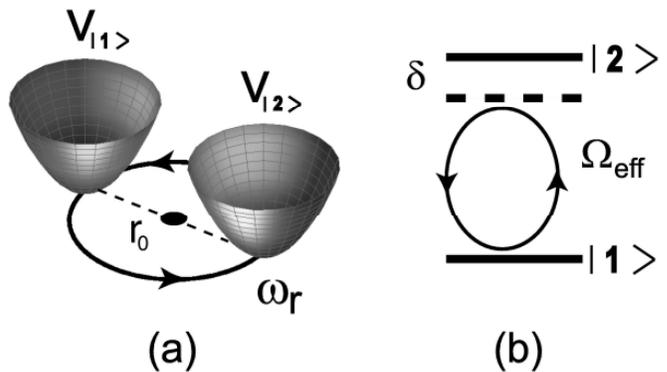}
\end{center}
\caption{Method for creating a vortex. (a) The two traps are rotated in the
  $xy$-plane about the $z$ axis at a frequency $\omega_r$. (b)
  Transitions are simultaneously driven between the two internal
  states at the effective Rabi frequency $\Omega_{\rm{eff}}$. A vortex
  mode possessing one unit of angular momentum can be prepared if
  $\omega_r \approx \Omega_{\rm eff}$.}
\end{figure}

We treat the internal structure of the condensed atoms as a two-level
system as illustrated in Fig.~1. The two states are confined in
separate axially symmetric harmonic oscillator potentials with the
same trap frequency $\omega_0$.  As shown in Fig.~1(a), the trap
centers are spatially offset by a distance $r_0$ and are rotated about
the symmetry axis at a frequency of $\omega_r$. Simultaneously an
electromagnetic field is applied that couples the two internal atomic
hyperfine states causing the atoms to coherently cycle between levels
as illustrated in Fig.~1(b). There are two parameters that
characterize the coupling: the detuning and the power. The detuning
$\delta$ denotes the mismatch of the frequency of the coupling
electromagnetic field to the frequency difference between the two
internal atomic states.  The power is characterized by the Rabi
frequency $\Omega$, which is the rate at which population would
oscillate between the two states if $\delta$ were zero. When $\delta$
is larger than $\Omega$, so that the drive is off resonance, the
population oscillations have small amplitude and occur at the
effective Rabi frequency $\Omega_{\rm eff}=\sqrt{\Omega^2+\delta^2}$.

The dynamical evolution of the condensate state $|\psi(t)\rangle$ for
a gas of atoms in two different hyperfine states is governed by a
nonlinear Schr\"{o}dinger equation, known as the Gross-Pitaevskii
equation, generalized to treat the internal state coupling$^5$
\begin{eqnarray}
  i \hbar {\partial\over\partial t} |\psi(t)\rangle &=&
  [\hat{H}_{0} \otimes \hat{1} + \hat{1}
  \otimes {\hbar\over2}(\Omega \,
  \hat{\sigma}_x + \delta \hat{\sigma}_z) \nonumber \\
  &&\quad{}+\hat{H}_1 \otimes \hat{\sigma}_z]
  |\psi(t)\rangle ,
\label{eq1}
\end{eqnarray}
where $\{\hat{1},\hat{\sigma}_x,\hat{\sigma}_y,\hat{\sigma}_z\}$ is
the set of standard Pauli spin operators and for clarity we have used
the tensor product $\otimes$ to explicitly separate the spatial and
and internal operators. The free Hamiltonian $\hat{H}_{0}$ describes
the motional dynamics of the atoms in a stationary harmonic trap
centered on the symmetry axis
\begin{equation}
  \hat{H}_{0}(\mbox{\boldmath{$r$}}) = -{\hbar^2\over{2m}} \nabla^2
  + \frac{1}{2} m \omega_0^2 r^2 + U_0 \, n(\mbox{\boldmath{$r$}}),
\label{eq2}
\end{equation}
where $m$ is the atomic mass. The term that makes the system nonlinear
is the mean-field interaction energy, which depends on the local
density of condensate atoms $n(\mbox{\boldmath{$r$}})$. The
coefficient $U_0=4\pi\hbar^2a/m$ is proportional to the scattering
length of a binary collision $a$. The effect of displacing the trap
centers and rotating them about the symmetry axis is described by
\begin{equation}
  \hat{H}_1(\mbox{\boldmath{$r$}},t) = \kappa[f(\mbox{\boldmath{$r$}})
  \cos(\omega_r t) + g(\mbox{\boldmath{$r$}}) \sin(\omega_r t)],
\label{eq3}
\end{equation}
where $\kappa$ is a coupling coefficient and
$f(\mbox{\boldmath{$r$}})$ and $g(\mbox{\boldmath{$r$}})$ are
functional prefactors. The explicit form of $\hat{H}_1$ determines the
symmetry of the quantum state being prepared. To create a vortex state
with one unit of angular momentum $\kappa=m\omega_0^2r_0$,
$f(\mbox{\boldmath{$r$}})=x$ and $g(\mbox{\boldmath{$r$}})=y$ in
Cartesian coordinates, corresponding to the scheme depicted in
Fig.~1(a). More general forms for $f$ and $g$ will be considered later
with Eq.~(3) then providing the definition of a generalized $\kappa$
and $\omega_r$.

Before solving this problem in detail, we first present an intuitive
picture of the underlying physics. The vortex state we wish to create
has unit circulation corresponding to a macroscopically occupied
single particle wave function with quantized angular momentum
$\langle\hat{L}_z\rangle/\hbar=1$.  In order to see why our scheme
couples a non-rotating condensate to this state, consider the frame
co-rotating with the trap centers at angular frequency $\omega_r$ so
that $\hat{H}_1$ becomes time-independent. The free Hamiltonian in the
co-rotating frame is given by$^{21}$ $\hat{H}_0-\omega_r\hat{L}_z$.
The energy of the vortex with one unit of angular momentum is
therefore shifted by $\hbar\omega_r$ in the rotating frame relative to
its value in the lab frame. When this energy shift compensates for
both the energy mismatch $\hbar\delta$ of the internal coupling field
and the small chemical potential difference between the vortex and
non-rotating condensate, resonant transfer of population may take
place.

For this picture to be valid, the following inequalities must be
satisfied
\begin{equation}
  \omega_r\gg\omega_0, \quad\delta\gg\Omega, \quad\omega_r\approx
  \Omega_{\rm eff},
\label{inequal}
\end{equation}
which also imply $\Omega_{\rm eff}\approx\delta$. The first inequality
splits significantly the energies of states of different angular
momenta in the rotating frame. This point is crucial for experimental
implementation since it allows the vortex to be generated rapidly and
is a feature absent from previous dynamic coupling schemes$^{18-20}$.
The weak coupling limit given by the second inequality allows the
resonance condition $\omega_r\approx\Omega_{\rm eff}$ to select
energetically the desired state with high fidelity.
\begin{figure}
\begin{center}\
\epsfysize=65mm
\epsfbox{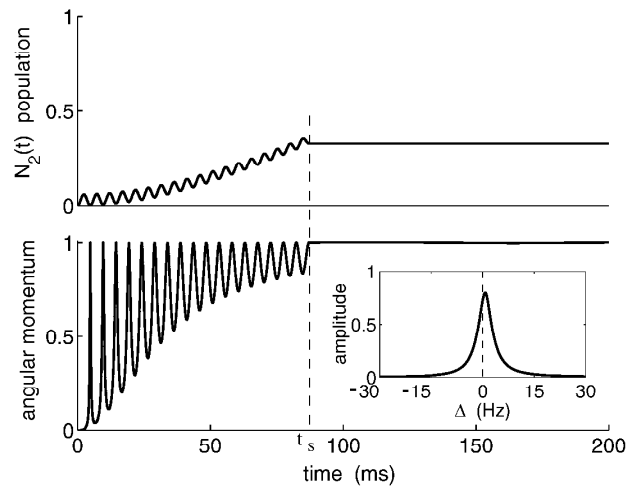}
\end{center}
\caption{Dynamical evolution to a vortex.  These are results of a numerical
  integration of Eq.~(\ref{eq1}), with the condensate initially in the
  non-rotating ground state and in the internal state $|1\rangle$. The
  coupling drive is turned on at time $t=0$, and is turned off at time
  $t=t_s$ by setting both $\Omega$ and $r_0$ to zero. The top graph
  shows the fractional population of atoms in the $|2\rangle$ internal
  state. The small-amplitude rapid oscillations correspond to the
  cycling between internal levels due to the off-resonant coupling.
  The gradual rise of this line is due to coupling from the ground
  state to the vortex mode caused by the drive $\hat{H}_1$ in
  Eq.~(\ref{eq3}). The bottom graph shows the angular momentum of the
  $|2\rangle$ state, in units of Planck's constant $\hbar$. The rise
  and fall of this curve corresponds to a rapid cycling of the
  $|2\rangle$ atoms between the non-rotating condensate and the
  vortex. Once during each Rabi cycle, the angular momentum approaches
  unity and at that time the $|2\rangle$ state wave function
  approaches a pure vortex mode. The inset shows the maximum amplitude
  of population transfer to the vortex as a function of the trap
  rotation frequency $\omega_r$, with
  $\Delta=\Omega_{\rm{eff}}-\omega_r$. The various parameters used in
  this calculation are: $\omega_0=10$ Hz, $\delta=200$ Hz,
  $\omega_r=205.4$ Hz, $N=8\times 10^5$ atoms, $m$ is the mass of the
  ${}^{87}$Rb atom, $a=5.5$ nm$^{23}$, and $\Omega=50$ Hz and $r_0=1.7
  \mu$m for $t<t_s$.}
\end{figure}
\noindent Even though we
require a large value for $\omega_r$, one may not use a simple
equivalent time-averaged potential for the effect of rotating the trap
centers without including the internal atomic state dynamics in the
coarse-graining, which occur on the same time scale.

In order to visualize the system dynamics, we present results from a
numerical integration of the Gross-Pitaevskii equation$^{22}$.  Here
we treat the system in two dimensions in the $xy$-plane perpendicular
to the symmetry axis. In Fig.~2 we plot the fractional population and
the angular momentum per atom of the $|2\rangle$ state as a function
of time. We initialize the system with all of the atoms in the
$|1\rangle$ internal level and in the mean-field ground state of the
trap. There are two striking features in Fig.~2.  Remarkably, even
though we are driving the internal states far from the atomic
resonance, a significant fraction of the population gets transferred
to the $|2\rangle$ state over a time long compared to the fast Rabi
oscillations at $\Omega_{\rm{eff}}$.  Furthermore, we see that the
angular momentum of state $|2\rangle$ oscillates rapidly at the
frequency $\Omega_{\rm{eff}}$.

\begin{figure}
\begin{center}\
\epsfysize=65mm
\epsfbox{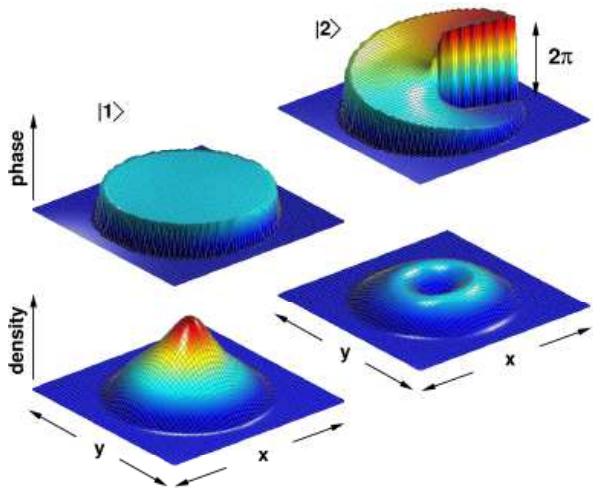}
\end{center}
\caption{Unit vortex preparation. Shown are the calculated densities and phases
  of the two states $|1\rangle$ and $|2\rangle$, at time
  $t=200\,\mbox{ms}$ indicated in Fig.~2. At this time, about one
  third of the atoms are in the $|2\rangle$ state, which is in a pure
  vortex mode with unit angular momentum.  A characteristic feature of
  a vortex mode is the $2n\pi$ phase wrap about the core, where $n$ is
  an integer that is equal to unity in this case. An attractive
  feature of this preparation scheme is that the $|1\rangle$ atoms
  residing in the core region provide a natural pinning mechanism for
  the vortex due to the mean-field repulsion.}
\end{figure}

Our key idea is that by turning off the coupling at a precise time,
$t=t_s$, on a given Rabi cycle, the $|2\rangle$ state can be prepared
to have unit angular momentum. The maximum possible population
transfer to the vortex state using this scheme obeys a Lorentzian
response curve as $\omega_r$ is varied near $\Omega_{\rm eff}$
exhibiting a narrow resonance. This is shown in the inset of Fig.~2,
where $\Delta = \Omega_{\rm{eff}}-\omega_r$, and will be discussed in
more detail later. In Fig.~3 we show a snapshot of the densities and
phases of the two components at time $t=200\,\mbox{ms}$.  The snapshot
illustrates the preparation of a high-quality vortex in state
$|2\rangle$, with the $|1\rangle$ state providing a natural
``pinning'' mechanism that stabilizes the vortex core by providing a
repulsive mean-field barrier along the symmetry axis. Consequently the
size of the vortex core is determined by the spatial density profile
of the non-rotating state, and not by the natural healing length.

An intriguing property of our novel state preparation scheme is that
the direction of circulation of the vortex can be opposite to that of
the rotating trap centers; changing the sign of the detuning $\delta$
with the direction of rotation of the trap centers fixed causes the
vortex to rotate in the opposite direction. This is most easily seen
by again considering the frame co-rotating with the potentials at
frequency $\omega_r$. Vortices with opposite circulations experience
opposite energy shifts in transforming to the rotating frame and
therefore require opposite signs of detuning in order to achieve
resonant coupling.

Although the numerical calculations we present here are full solutions
of Eq.~(\ref{eq1}), an accurate and greatly simplified model may be
derived using the separation of time scales given by the inequalities
in Eq.~(\ref{inequal}). This allows us to gain insight into the
physical mechanisms giving rise to the behavior of this system. If we
coarse-grain over rapidly oscillating terms in order to study the long
time scale behavior, we arrive at an approximate solution for the
system dynamics, which has the simple form$^{24}$
\begin{eqnarray}
 |\psi(t)\rangle &=& [a(t) c_0(t)\,|\phi_0\rangle
 + b(t) c_n(t)\,|\phi_n\rangle] |1\rangle \nonumber \\
&+& [b(t) c_0(t)\,|\phi_0\rangle
 + a^*(t) c_n(t)\,|\phi_n\rangle] |2\rangle. 
\label{eq4}
\end{eqnarray}
Here $|\phi_0\rangle$ and $|\phi_n\rangle$ are the spatial parts of
the state, with $|\phi_0\rangle$ being the ground state and
$|\phi_n\rangle$ being the target state---the unit vortex for the case
considered in Fig.~3, determined by the self-consistent lowest energy
solutions of the two-component Gross-Pitaevskii equation$^5$. For
given populations of the two atomic states, the energy is minimized
using a steepest descents algorithm with the constraint that
$|\phi_0\rangle$ is the nodeless ground state and $|\phi_n\rangle$
possesses the symmetry imposed by $\hat{H}_{1}$, which for the unit
vortex is a $2\pi$ topological phase wrap on any closed loop
containing the core. The rapidly varying coefficients in
Eq.~(\ref{eq4}) are found to be given by $a(t) =
\cos(\Omega_{\rm{eff}} t/2) - i(\delta / \Omega_{\rm{eff}})
\sin(\Omega_{\rm{eff}} t/2)$ and $b(t) = -i (\Omega /
\Omega_{\rm{eff}})\sin(\Omega_{\rm{eff}}t/2)$. The slow coupling of
the spatial states is given by $c_0$ and $c_n$, which satisfy
\begin{equation}
\begin{array}{ccc}
i  \hbar \left(\begin{array}{c}  
       \dot{c}_0  \\ \dot{c}_n  \end{array}\right)
     &=& {1\over2} \left(\begin{array}{cc}  
    -(\epsilon_n-\epsilon_0-\hbar\Delta)  & \beta
                             \langle\phi_n| f+ig| \phi_0\rangle\\
     \beta
     \langle\phi_0| f-ig| \phi_n\rangle
                             & (\epsilon_n-\epsilon_0-\hbar\Delta)
\end{array}   \right)    \left(   \begin{array}{c}  
       c_0  \\  c_n  \end{array}   \right)
\end{array} .
\label{eq5}
\end{equation}
Here $\beta= \kappa \Omega/|\delta|$ is the coupling coefficient, $f$
and $g$ are defined in Eq.~(\ref{eq3}), and $\epsilon_0$ and
$\epsilon_n$ are the energy eigenvalues of $|\phi_0\rangle$ and
$|\phi_n\rangle$, respectively. The time scale for state preparation
is determined by $\beta$ and the overlap matrix element
$\langle\phi_n| f+ig| \phi_0\rangle$. These two parameters also
determine the width of the response Lorentzian that is shown in the
inset of Fig.~2 with the maximum occurring not at zero frequency but
displaced by an amount according to the splitting between the
eigenenergies $\epsilon_n-\epsilon_0$; typically a small fraction of
$\omega_0$.

The symmetry of the coupling is determined by $f+ig$, which has the
form $x+iy$ as previously stated for the case of the unit vortex. In
order to produce a vortex with $n$ units of angular momentum (i.e.
$\langle\hat{L}_z\rangle/\hbar=n$), $f+ig$ should take on the form
$(x+iy)^n$.  So for $n=2$, we must construct an $\hat{H}_1$ in which
$f=x^2-y^2$ and $g=2xy$, corresponding to a saddle potential rotating
at $\omega_r/2$. In Fig.~4, we illustrate vortex generation with two
and three units of angular momentum.

\begin{figure}
\begin{center}\
\epsfysize=65mm
\epsfbox{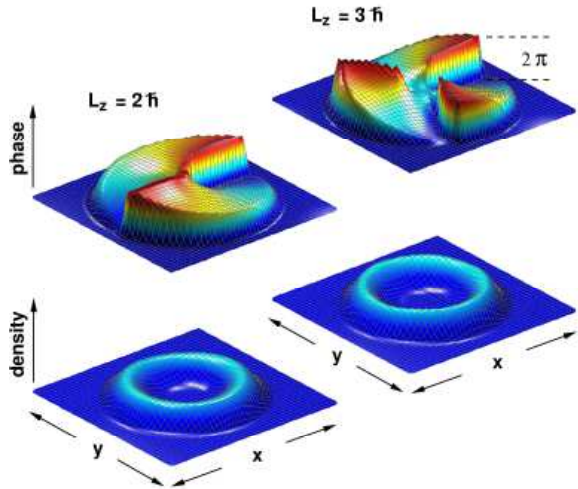}
\end{center}
\caption{Double and triple vortex preparation. Shown are the densities and
  phases modulo $2\pi$ of the $|2\rangle$ state after a dynamical
  evolution similar to that shown in Fig.~2, but with different
  symmetries of the drive $\hat{H}_1$.  In the case of
  $\langle\hat{L}_z\rangle/\hbar=2$, $f=x^2-y^2$ and $g=2xy$, while
  for $\langle\hat{L}_z\rangle/\hbar=3$, $f=x^3-3xy^2$ and
  $g=3x^2y-y^3$.  In both cases, the system was evolved from the same
  initial condition as that for the calculation described in Fig.~2,
  with about one third of the atoms in the $|2\rangle$ state at the
  time $t_s$.  The values taken for the various parameters were the
  same, except for the trap rotation frequency, which was
  $\omega_r=204.3$ Hz for the $\langle\hat{L}_z\rangle/\hbar=2$ case,
  and $\omega_r=200.2$ Hz for the $\langle\hat{L}_z\rangle/\hbar=3$
  case.}
\end{figure}

The generalization of our scheme for the preparation of macroscopic
quantum states of arbitrary symmetry is straightforward. For example,
to generate a mode with a dipole symmetry one superimposes a left and
right circulating vortex so that $f={\rm{Re}}\{x+iy\}$ and $g=0$. A
quadrapole symmetry would be generated by $f={\rm{Im}}\{(x+iy)^2\}$
and $g=0$. The dynamical state preparation of these two examples is
illustrated in Fig.~5.

We do not anticipate collisions to play a significant role on the
short time scale for vortex generation shown in Fig.~2 and we have
implicitly neglected them in our zero temperature theory. However, the
system is thermodynamically unstable on long time scales if the
rotating drive $\hat{H}_1$ is left on, since atoms can access states
of higher and higher angular momentum through collisional relaxation.
Even after the vortex state has been generated and the drive is turned
off, collisions will be important in determining the stability of
persistent currents where the topological structure of the two
component order parameter is known to be crucial$^{3,12}$.

\begin{figure}
\begin{center}\
\epsfysize=65mm
\epsfbox{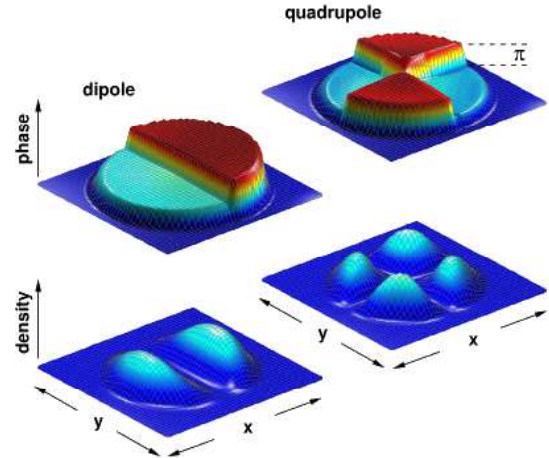}
\end{center}
\caption{Dipole and quadrupole preparation. Different symmetries were
  constructed using specific forms for the drive $\hat{H}_1$ in order
  to prepare the $|2\rangle$ state in non-circulating modes. Instead
  of having a $2n\pi$ phase wrap corresponding to a current flowing
  around a central core, these modes have regions of constant phase
  separated by a discontinuous jump of $\pi$ where the wave function
  changes sign.  To generate the dipole mode, we used $f=x$ and $g=0$
  with $\omega_r=205.4$ Hz; while for the quadrupole, we used $f=xy$
  and $g=0$ with $\omega_r=204.3$ Hz.}
\end{figure}

{\em Note added:} While this paper was in press, vortices in a
two-component Bose-Einstein condensate have been created and observed
using this scheme$^{25}$.

\vskip1pc

\noindent
\newlength{\colwidth}
\setlength{\colwidth}{8.7cm}
\rule[.03in]{\colwidth}{.005in}

\newlength{\cola}
\settowidth{\cola}{25}
\addtolength{\cola}{1em}
\newlength{\colb}
\setlength{\colb}{\colwidth}
\addtolength{\colb}{-\cola}

\vfill

\noindent\begin{tabular}{p{\cola}p{\colb}}

1.& Cornell, E. A., Ensher, J. R. \& Wieman, C. E. Experiments in
Dilute Atomic Bose-Einstein Condensation. {\em Proceedings of the 1998
  Enrico Fermi summer school on Bose-Einstein condensation in Varenna,
  Italy.} Preprint http://xxx.lanl.gov {\em cond-mat/9903109}.\\
  
2.& Ketterle, W., Durfee, D. S. \& Stamper-Kurn, D. M.  Making,
probing and understanding Bose-Einstein condensates. {\em Proceedings
  of the 1998 Enrico Fermi summer school on Bose-Einstein condensation
  in Varenna, Italy,} {\em cond-mat/9904034}.\\
  
3.& Leggett, A. Low temperature physics, superconductivity and
superfluidity. In {\em The New Physics}. Editor Davies, P., Cambridge
University Press, London, 268--288 (1989).\\
  
\end{tabular}  
\noindent\begin{tabular}{p{\cola}p{\colb}}

4.& Matthews, M. R., Anderson, B. P., Haljan, P. C., Hall, D. S.,
Williams, J. E., Holland, M. J., Wieman, C. E. \& Cornell, E. A.
Watching a superfluid untwist itself: Recurrence of Rabi oscillations
in a Bose-Einstein condensate, {\em cond-mat/9906288}.\\
  
5.& Williams, J., Walser, R., Cooper, J., Cornell, E. A. \& Holland,
M. Excitation of an antisymmetric collective mode in a strongly
coupled two-component Bose-Einstein condensate, {\em
  cond-mat/9904399}.\\
  
6.& Anderson, M. H., Ensher, J. R., Matthews, M. R., Wieman, C. E. \&
Cornell, E. A. Observation of Bose-Einstein condensation in a dilute
atomic vapor. {\em Science} {\bf 269}, 198--201 (1995).\\

7.& Davis, K. B., Mewes, M.-O., Andrews, M. R., van Druten, N. J.,
Durfee, D. S., Kurn, D. M. \& Ketterle, W. Bose-Einstein condensation
in a gas of sodium atoms.  {\em Phys. Rev.  Lett.} {\bf 75},
3969--3973 (1995).\\
  
8.& Bradley, C. C., Sackett, C. A., Tollet, J. J. \& Hulet, R. G.
Evidence of Bose-Einstein Condensation in an Atomic Gas with
Attractive Interactions. {\em Phys.\ Rev.\ Lett.} {\bf 75}, 1687--1690
(1995); Erratum. {\em Phys. Rev.  Lett.} {\bf 79}, 1170--1170 (1997).\\
  
9.& Butts, D. A. \& Rokhsar, D. S.  Predicting signatures of rotating
Bose-Einstein condensates. {\em Nature} {\bf 397}, 327--329 (1999).\\
  
10.& Rokhsar, D. S. Vortex stability and persistent currents in
trapped Bose gases. {\em Phys.\ Rev.\ Lett.} {\bf 79}, 2164--2167
(1997).\\
  
11.& Dalfovo, F. \& Stringari, S.  Bosons in anisotropic traps: Ground
state and vortices. {\em Phys.\ Rev.\ A} {\bf 53}, 2477--2485 (1996).\\
  
12.& Ho, T.-L. \& Shenoy, V. B. Local spin-gauge symmetry of the
Bose-Einstein condensates in atomic gases. {\em Phys.\ Rev.\ Lett.}
{\bf 77}, 2595--2599 (1996).\\
  
13.& Barenghi, C. F. Vortex waves in a cloud of
Bose-Einstein-condensed, trapped alkali-metal atoms. {\em Phys.\ Rev.\ 
  A} {\bf 54}, 5445--5446 (1996).\\
  
14.& Fetter, A. L. Vortex stability in a trapped Bose condensate. {\em
  J. Low Temp.\ Phys.} {\bf 113}, 189--194 (1998).\\
  
15.& Feder, D. L., Clark, C. W. \& Schneider, B. I. Vortex stability
of interacting Bose-Einstein condensates confined in anisotropic
harmonic traps. {\em Phys.\ Rev.\ Lett.} {\bf 82}, 4956--4959 (1999).\\
  
\end{tabular}  
\noindent\begin{tabular}{p{\cola}p{\colb}}

16.& Caradoc-Davies, B. M., Ballagh, R. J. \& Burnett, K. Coherent
dynamics of vortex formation in trapped Bose-Einstein condensates.
{\em Phys.\ Rev.\ Lett.} {\bf 83}, 895--898 (1999).\\
  
17.& Dodd, R. J., Burnett, K., Edwards, M.  \& Clark, C. W. Excitation
spectroscopy of vortex states in dilute Bose-Einstein condensed gases.
{\em Phys.\ Rev.\ A} {\bf 56}, 587--590 (1997).\\
  
18.& Marzlin, K.-P. \& Zhang, W.  Vortex coupler for atomic
Bose-Einstein condensates. {\em Phys.\ Rev.\ Lett.} {\bf 79},
4728--4731 (1997).\\
  
19.& Bolda, E. L. \& Walls, D. F.  Creation of vortices in a
Bose-Einstein condensate by a Raman technique. {\em Phys.\ Lett.\ A}
{\bf 246}, 32--36 (1998).\\
  
20.& Dum, R., Cirac, J. I., Lewenstein, M. \& Zoller, P.  Creation of
dark solitons and vortices in Bose-Einstein condensates. {\em Phys.\ 
  Rev.\ Lett.} {\bf 80}, 2972--2975 (1998).\\
  
21.& Lifshitz, E. M. \& Pitaevskii, L.
P. {\em Statistical Physics} Pergamon, Oxford (1980).\\
  
22.& Holland, M. J., Jin, D. S., Chiofalo, M. L. \& Cooper, J.
Emergence of interaction effects in Bose-Einstein condensation. {\em
  Phys.\ Rev.\ Lett.} {\bf 78}, 3801--3805 (1997).\\
  
23.& Hall, D. S., Matthews, M. R., Ensher, J. R., Wieman, C. E. \&
Cornell, E. A.  Dynamics of component separation in a binary mixture
of Bose-Einstein condensates. {\em Phys.\ Rev.\ Lett.} {\bf 81},
1539--1542 (1998).\\
  
24.& Williams, J. E. The Preparation of Topological Modes in a
Strongly-Coupled Two-Component Bose-Einstein Condensate. {\em Ph.D.
  Thesis, University of Colorado at Boulder} (1999).\\
  
25.& Matthews, M. R., Anderson, B. P., Haljan, P. C., Hall, D. S.,
Wieman, C. E. \& Cornell, E. A.  Vortices in a Bose-Einstein
Condensate, {\em cond-mat/9908209}.
\end{tabular}

\vskip3pc


{\bf Acknowledgements.} We would like to thank E. A.  Cornell, M. R.
Matthews, P. C. Haljan, B. P.  Anderson, and C. E. Wieman, for
extensive discussions on the realization of our scheme. We would also
like to thank R. Walser and J. Cooper for helpful comments.  This work
was supported by the National Science Foundation.

\end{document}